\documentclass[a4paper,12pt]{article}
\usepackage{amsmath}
\usepackage{amssymb}
\usepackage{latexsym}
\newcommand{\be}{\begin{eqnarray}}
\newcommand{\ee}{\end{eqnarray}}
\begin{document}
\title{\large\textbf{Space Time Covariance of Canonical Quantization of Gravity: A
(formal) general result and the (rigorous) explicit case of 2+1
Quantum Cosmology}}
\author{\textbf{T. Christodoulakis}\thanks{tchris@cc.uoa.gr} ~~~\textbf{G.O. Papadopoulos}\thanks{gpapado@cc.uoa.gr}\\ University of Athens, Physics Department\\
Nuclear \& Particle Physics Section\\
Panepistimioupolis, Ilisia GR 157--71, Athens, Hellas}
\date{}
\maketitle
\begin{center}
\textit{}
\end{center}
\vspace{0.5cm} \numberwithin{equation}{section}
\begin{abstract}
A general classical theorem is presented according to which all
invariant relations among the space time metric scalars, when
turned into functions on the Phase Space of full Pure Gravity
(using the Canonical Equations of motion), become weakly vanishing
functions of the Quadratic and Linear Constraints. The implication
of this result is that (formal) Dirac consistency of the Quantum
Operator Constraints (annihilating the wave Function) suffices to
guarantee space time covariance of the ensuing quantum theory: An
ordering for each invariant relation will always exist such that
the emanating operator has an eigenvalue identical to the
classical value. \\ The example of 2+1 Quantum Cosmology is
explicitly considered: The four possible ``Cosmological
Solutions'' --two for pure Einstein's equations plus two more when
a $\Lambda$ term is present- are exhibited and the corresponding
models are quantized. The invariant relations describing the
geometries are explicitly calculated and promoted to operators
whose eigenvalues are their corresponding classical values.
\end{abstract}
\newpage
\section{Introduction}
The problem of space time covariance of a Quantum Theory of
Gravity \cite{Isham} within the context of Canonical Quantization
can essentially be described as follows: Classically, Einstein's
Field equations are known to be (not manifestly but explicitly)
equivalent to the Hamiltonian and Momentum constraints plus the
Canonical Equations of motion. This is understandable since,
although the canonical analysis uses objects defined on the
hypersurface, the momenta involve the extrinsic curvature and thus
carry the information of the embedding of the hypersurface in
space time. When we canonically quantize, the momenta become
functional derivatives with respect to the spatial metric (choice
of Polarization in the phase space) and thus the fate of space
time covariance is somewhat obscure. Of course if one takes
special care so that the geometrical meaning of the classical
constraints is maintained at the quantum level, one is justified
to expect space time covariance of the ensuing theory. Indeed as
we shall see in the first section, when we suitably define the
equivalence between two sets of state vectors emerging upon
quantizing the same space time in two different foliations, then
space time covariance is indeed achieved.

In the next sections we present the example of 2+1 spatially
homogeneous cosmological models in the absence as well as in the
presence of a $\Lambda$ term. The classical solutions i.e. the
different ``cosmological'' disguises of Minkowski space time or
the maximally symmetric space are explicitly given. The
corresponding models are quantized and the classical observables
are turned into operators in such a way that their eigenvalues
coincide with the classical values. In this sense it is proven
that the two pairs of state vectors are equivalent.

\section{A Classical Theorem and its Quantum Implications}
As is well known \cite{Sundermeyer}, the canonical analysis of
pure Gravity consists in the following statements:
\begin{subequations}
\be
\mathcal{H}_{0}(g_{ij},\pi^{ij})\approx 0\\
\mathcal{H}_{k}(g_{ij},\pi^{ij})\approx 0\\
\dot{g}_{ij}=\{g_{ij},H\}\\
\dot{\pi}^{ij}=\{\pi^{ij},H\}\\
H=\int(N^{0}\mathcal{H}_{0}+N^{k}\mathcal{H}_{k})d^{3}x
\ee
\end{subequations}
which are explicitly equivalent to the ten Einstein's Field
Equations. If we adopt the notion that classical observables, are
all the geometrical objects that do not depend on the gauge, i.e.
the coordinate system, then we are led to identify these
observables with invariant relations among space time scalars.
These scalars can be constructed in two ways: Firstly, by
contracting all the indices of tensor products of the Riemann
tensor and its covariant derivatives of any order.Secondly,in the
case of spacetimes admitting a null, covariantly constant vector
field ( pp waves where all the scalars constructed in the above
mentioned way are identically vanishing ), by finding
proportionality factors between tensors constructed by the Riemann
tensor and its covariant derivatives of any order. Anyway, the
scalars themselves do not describe the space time in an invariant
manner since their functional form in terms of the coordinates
changes when the coordinate system is altered. A way to generate
invariant relations among these scalars, albeit not the most
efficient one, would be to take a base of 4 scalars (say $Q_{1},
\ldots Q_{4}$) and solve for the coordinates. Then, any other
scalar (say $Q_{5}$) becomes expressible in terms of the 4 scalars
chosen ($Q_{5}=f(Q_{1},\ldots,Q_{4})$). Now this relation is
characteristic of the geometry i.e. does not change form under
coordinate transformations. In this sense a geometry is complectly
characterized by a set of relations: \be f^{A}(Q_{i})=0 \ee The
index $A$ is at most countable. Turning these relations into
functions on the phase space we notice that they become weakly
vanishing quantities: \be h^{A}(g_{ij},\pi^{ij})\approx 0 \ee In
implementing this step use has been made of canonical equations of
motion in order to substitute all higher time derivatives of the
metric of the slice. But at this point, we invoke a known theorem
of constrained dynamics \cite{Sundermeyer}:
\paragraph{Theorem}
``Every weakly vanishing function in Phase Space is strongly equal to some
expression containing the Constraints (which define a surface in Phase Space)''\\
Moreover the expression under discussion ought to vanish on-mass shell (i.e. when the
constraints are set to zero). Thus:
\be
h^{A}=h^{A}(\mathcal{H}_{0},\mathcal{H}_{k})
\ee

The translation of the above result in the velocity phase space reads as follows:
Consider an invariant relation $ f^{A}(Q_{i})=0 $ of any space time geometry which satisfies Einstein's
Field Equations. Evaluate the left-hand side for the generic space time metric. Then, eliminate from the resulting expression
all higher time derivatives of the metric ( on the slice ) using the spatial Einstein equations ( and their time derivatives of the appropriate order) .
The end result will be that $f^{A}$  will become such a function of the $G^{0}_{0}, G^{0}_{k}$
constraints so that it vanishes when the constraints are set to zero . The proof rests on the fact that the constraints are the only quantities
of the spatial metric and its time derivative that vanish, a thing that is guaranteed by the consistency of the Field Equations;
Any other function on the velocity Phase Space which vanishes by virtue of the Field Equations, such as $f^{A}$,
is necessarily expressible in terms of the constraints.

The implications of this for the quantum theory are obvious and
important: Adopting the point of view that the quantum observables
are to be the operator analogues of the classical observables, we
are assured that,for each and every such observable, there will
exist a factor ordering such that all ensuing operators, when
acting on the appropriate states defined by:
\begin{subequations} \label{relationsuponPsi}
\be
\mathcal{\widehat{H}}_{0}\Psi=0\\
\mathcal{\widehat{H}}_{k}\Psi=0
\ee
\end{subequations}
will have vanishing eigenvalues:
\be
\widehat{h}^{A}\Psi=0
\ee
This establishes the (formal) space time covariance of the quantum theory described by
(\ref{relationsuponPsi}) above.

\section{Preliminaries on Spatially Homogeneous Spacetimes}
We are now interested in exemplifying the theorem and its
implications by means of a concrete, finite-dimensional Quantum
Cosmology example. To this end, we wish to manufacture a situation
in which a specific space time admits two different homogeneous
foliations. This is generically impossible in 3+1 Cosmology, where
the imposition, on the hypersurface, of the various 3-dim Lie
Groups as symmetry groups of motion, leads to the distinct Bianchi
models. The very rare exceptions have to do with limiting cases,
such as Milne's solution \cite{Milne} which is 4-dim Minkowski
geometry disguised as a Type V cosmology, or the trivial case of
Kasner's universes with two of the three scale factors constant. A
suitable situation that comes to mind is the case of 2+1
dimensions: There, the pure Einstein's Field Equations admit only
the Minkowski geometry as a solution, while when a $\Lambda$ term
is present, the only possibility is a maximally symmetric space.
Thus, if we solve the general Field Equations under the
restriction of spatial homogeneity, we are bound to end up with
``cosmological'' descriptions of either Minkowski geometry or a
space of maximal symmetry , correspondingly.

Before proceeding with the examples, it is deemed as appropriate to exhibit some
basic assumptions.

Spacetime, is assumed to be the pair $(\mathcal{M}, g)$ where
$\mathcal{M}$ is a 3-dimensional, Hausdorff, connected,
time-oriented and $C^{\infty}$ manifold, and g is a $(0,2)$ tensor
field, globally defined, $C^{\infty}$, non degenerate and
Lorentzian i.e. it has signature $(-,+,+)$. In the spirit of 2+1
analysis, $\mathcal{M}=R\times \Sigma_{t}$, where the
2-dimensional orientable submanifolds $\Sigma_{t}$ (\emph{surfaces
of simultaneity}), are spacelike surfaces of constant time; their
evolution in time, results in the entire spacetime. The
assumption of spatial homogeneity  corresponds to the
imposition of the action of a symmetry group of transformations
$G$ upon the manifolds $\Sigma_{t}$. Usually the group $G$ is not
only continuous, but also a Lie group --thus denoted by $G_{r}$,
where $r$ is the dimension of the space of its parameters.
Avoiding the details on these issues --these matters can be easily
found in every standard reference see e.g.
\cite{Eisenhart-Petrov}-- we simply state that spatially
homogeneous models with a simply transitive action of the symmetry group
 are described (apart the topology of $\Sigma_{t}$) by an invariant basis of one forms
$\sigma^{\alpha}_{i}(x)dx^{i}$. By invariant we mean that  their Lie Derivative with
respect to the generators of the Lie Group $G_{r}$ are zero. The case of Groups which act
multiply transitively will not concern us. \\
The line element of such a space time can be casted into the form:
\be \label{lineelement}
ds^{2}=(N^{\alpha}(t)N_{\alpha}(t)-N^{2}(t))dt^{2}+2N_{\alpha}(t)\sigma^{\alpha}_{i}(x)
dtdx^{i}+\gamma_{\alpha\beta}(t)\sigma^{\alpha}_{i}(x)\sigma^{\beta}_{j}(x)dx^{i}dx^{j}
\ee with: \be \label{rotation}
\sigma^{\alpha}_{i,j}(x)-\sigma^{\alpha}_{j,i}(x)=2C^{\alpha}_{\mu\nu}
\sigma^{\nu}_{i}(x)\sigma^{\mu}_{j}(x) \ee where
$\gamma_{\alpha\beta}(t)$ is the metric induced on the surfaces
$\Sigma_{t}$ (and constant on them), $N(t)$ is the lapse function,
$N_{\alpha}(t)$ is the shift vector
($N^{\alpha}(t)=\gamma^{\alpha\beta}(t)N_{\beta}(t)$,
$\gamma^{\alpha\beta}(t)$ being the inverse of
$\gamma_{\alpha\beta}(t)$), and $C^{\alpha}_{\mu\nu}$ are the
structure constants for the corresponding Lie Algebra. Greek
indices count the different one-forms, while the Latin  are world
indices; both kind of indices range in the interval [1,2] . In 2
dimensions,there are only two distinct Lie Algebras; the
Abelian(I) , where all the structure constant vanish
$C^{\alpha}_{\mu\nu}=0$ and the Non Abelian (II), where there is
only one independent  non vanishing structure constant --say
$C^{1}_{12}=1$; every other choice for the non vanishing structure
constants can be transformed into this, under a linear mixing of
the initial set i.e. using a new set of the form
$\widetilde{C}^{\alpha}_{\mu\nu}=\Upsilon^{\alpha}_{\beta}\Delta^{\kappa}_{\mu}
\Delta^{\lambda}_{\nu} C^{\beta}_{\kappa\lambda}$, where
$\Delta^{\alpha}_{\beta}$ $\in GL(2,\mathbb{R})$, and
$\Upsilon^{\alpha}_{\beta}$ its inverse. Thus, two
``cosmological'' models emerge for the pure gravity case, and two
more when a $\Lambda$ term is present.

At this point, a question arises; is there any particular class of
General Coordinate Transformations (G.C.T.'s) which can serve to
simplify the form of Einstein's Field Equations (E.F.E.'s)? The
answer is positive and a thorough investigation of this problem
and its consequences, is given in \cite{JMP}; indeed, not only
there is a class of G.C.T.'s which preserve the manifest spatial
homogeneity of the line element (\ref{lineelement}), but also
forms a continuous (and virtually, Lie) group. This group is
closely related to the symmetries of the
symmetry Lie Group $G_{r}$; it is its automorphism group.
A brief account of the relevant findings of \cite{JMP} is as follows:

Consider the transformations
\begin{subequations}\label{transformations}
\begin{align}
t\rightarrow \widetilde{t}=t&\Leftrightarrow t=\widetilde{t}\\
x^{i}\rightarrow \widetilde{x}^{i}=g^{i}(t,x^{j})&\Leftrightarrow
x^{i}=f^{i}(t,\widetilde{x}^{j})
\end{align}
\end{subequations}
When we insert (\ref{transformations})  into (\ref{lineelement}), the need to preserve the manifest spatial
homogeneity of the latter, leads to the restrictions:
\begin{subequations}\label{partialrelations}
\begin{align}
\frac{\partial f^{i}}{\partial t}&=\sigma^{i}_{\alpha}(f)P^{\alpha}(t)\\
\frac{\partial f^{i}}{\partial
\widetilde{x}^{j}}&=\sigma^{i}_{\alpha}(f)
\Lambda^{\alpha}_{\beta}(t)
\sigma^{\beta}_{j}(\widetilde{x})
\end{align}
\end{subequations}
and,subsequently, to the identifications :
\begin{subequations}\label{gaugetransfor}
\begin{align}
\widetilde{N}(t)&=N(t)\\
\widetilde{N}^{\alpha}(t)&=S^{\alpha}_{\beta}(t)(N^{\beta}(t)+
P^{\beta}(t))\\
\widetilde{\gamma}_{\alpha\beta}(t)&=\Lambda^{\mu}_{\alpha}(t)
\Lambda^{\nu}_{\beta}(t) \gamma_{\mu\nu}(t)
\end{align}
\end{subequations}
where $N^{\alpha}(t)=\gamma^{\alpha\beta}(t)N_{\beta}(t,)$ , with
$\sigma^{i}_{\alpha}(x)$ being the matrix inverse to
$\sigma^{\alpha}_{i}(x)$ Integrability conditions for the system
(\ref{partialrelations})  i.e. Frobenious' Theorem, results in the
system (the dot, whenever used, denotes differentiation with
respect to time):
\begin{subequations} \label{Aut}
\begin{align}
C^{\beta}_{\mu\nu}\Lambda^{\alpha}_{\beta}(t)&=C^{\alpha}_{\kappa\lambda}
\Lambda^{\kappa}_{\mu}(t)\Lambda^{\lambda}_{\nu}(t)\\
\frac{1}{2}\dot{\Lambda}^{\alpha}_{\beta}(t)&=C^{\alpha}_{\mu\nu}P^{\mu}(t)
\Lambda^{\nu}_{\beta}(t)
\end{align}
\end{subequations}
and, \emph{``Time-Dependent Automorphisms Inducing
Diffeomorphisms''} emerge. The automorphisms of a Lie Group
$G_{r}$ constitute a continuous group. The members of the group
which are continuously connected to the identity element, form a
Lie Group as well --though the topology of the latter might be
different from that of the former. If one considers parametric
families of the automorphic matrices, characterized by the
parameters $\tau^{i}$, $\Lambda^{\alpha}_{\beta}(t;\tau^{i})$ and,
as usual, demands:
\begin{subequations}
\begin{align}
\Lambda^{\alpha}_{\beta}(t;\tau^{i})\Big|_{\tau^{i}=0}&=\delta^{\alpha}_{\beta}\\
\frac{d\Lambda^{\alpha}_{\beta}(t;\tau^{i})}{d\tau^{i}}\Big|_{\tau^{j\neq
i}=0}&= \lambda^{\alpha}_{\beta(i)}
\end{align}
\end{subequations}
where $\lambda^{\alpha}_{\beta(i)}$, are the generators with
respect to the parameter $\tau^{i}$, of the Lie Algebra of the
Automorphisms, then from the first of (\ref{Aut}), after a
differentiation with respect to $\tau^{i}$, will have:
\be\label{generators}
\lambda^{\alpha}_{\beta(i)}C^{\beta}_{\mu\nu}=\lambda^{\rho}_{\mu(i)}
C^{\alpha}_{\rho\nu}+\lambda^{\rho}_{\nu(i)}C^{\alpha}_{\mu\rho}
\ee
For an extensive treatment on these issues see \cite{CMP},
while for the relation and usage of these generators with
conditional symmetries \cite{Kuchar} see \cite{CQGTYPEI}.

In n+1 analysis (here n=2), the E.F.E.'s in vacuum, when the
cosmological constant $\Lambda_{c}$ exists, take the form:
\begin{subequations} \label{EinsteinEq}
\begin{align}
E^{0}_{0}&=K^{\alpha}_{\beta}K^{\beta}_{\alpha}-K^{2}+(R+2\Lambda_{c})=0
\label{quadraticconstraint}\\
E^{0}_{\alpha}&=K^{\mu}_{\nu}C^{\nu}_{\alpha\mu}-K^{\mu}_{\alpha}
C^{\nu}_{\mu\nu}=0\label{linearconstraint}\\
E^{\alpha}_{\beta}&=\dot{K}^{\alpha}_{\beta}-NKK^{\alpha}_{\beta}+N(R^{\alpha}_{\beta}+
\frac{2}{n-1}\Lambda_{c}\delta^{\alpha}_{\beta})+
2N^{\rho}(K^{\alpha}_{\nu}C^{\nu}_{\beta\rho}-K^{\nu}_{\beta}C^{\alpha}_{\nu\rho})=0
\label{equationofmotion}
\end{align}
\end{subequations}
with:
\begin{subequations}
\begin{align}
K^{\alpha}_{\beta}(t)&=\gamma^{\alpha\rho}(t)K_{\rho\beta}(t)\\
R^{\alpha}_{\beta}(t)&=\gamma^{\alpha\rho}(t)R_{\rho\beta}(t)\\
K_{\alpha\beta}(t)&=-\frac{1}{2N(t)}(\dot{\gamma}_{\alpha\beta}(t)+
2\gamma_{\alpha\nu}(t)
C^{\nu}_{\beta\rho}N^{\rho}(t)+2\gamma_{\beta\nu}(t)C^{\nu}_{\alpha\rho}N^{\rho}(t))\\
\begin{split} \label{curvature}
R_{\alpha\beta}(t)&=C^{\kappa}_{\sigma\tau}
C^{\lambda}_{\mu\nu}\gamma_{\alpha\kappa}(t)
\gamma_{\beta\lambda}(t)\gamma^{\sigma\nu}(t)\gamma^{\tau\mu}(t)+
2C^{\lambda}_{\alpha\kappa}C^{\kappa}_{\beta\lambda}+
2C^{\mu}_{\alpha\kappa}C^{\nu}_{\beta\lambda}\gamma_{\mu\nu}(t)
\gamma^{\kappa\lambda}(t)\\
&+2C^{\lambda}_{\alpha\kappa}C^{\mu}_{\mu\nu}\gamma_{\beta\lambda}(t)
\gamma^{\kappa\nu}(t)+
2C^{\lambda}_{\beta\kappa}C^{\mu}_{\mu\nu}\gamma_{\alpha\lambda}(t)
\gamma^{\kappa\nu}(t)
\end{split}
\end{align}
\end{subequations}
where, $\delta^{\alpha}_{\beta}$ is the Kronecker's Delta. \\ Since
G.C.T.'s , in general, are covariances of the E.F.E.'s, the same must hold true for the
Time Dependent A.I.D.'s. Indeed, under the ``gauge'' transformations (\ref{gaugetransfor}) one can readily proof
(with the help of $\ref{Aut}$) :
\begin{subequations}
\begin{align}
\widetilde{E}^{0}_{0}&=E^{0}_{0}\\
\widetilde{E}^{0}_{\alpha}&=\Lambda^{\beta}_{\alpha}E^{0}_{\beta}\\
\widetilde{E}^{\alpha}_{\beta}&=S^{\alpha}_{\kappa}\Lambda^{\lambda}_{\beta}
E^{\kappa}_{\lambda}
\end{align}
\end{subequations}
where $S^{\alpha}_{\beta}$ is the matrix inverse to
$\Lambda^{\alpha}_{\beta}$. The effect of a time
reparameterization can also be trivially seen to be a covariance
of (\ref{EinsteinEq}). Using the freedom contained in
$P^{\alpha}(t)$ one can always set the shift to zero
which,together with the choice of time $N(t)=\sqrt{\gamma(t)}$
considerably simplifies the E.F.E.'s (\ref{EinsteinEq}) which
become:
\begin{subequations}
\begin{align}
&-\gamma^{\alpha\mu}\gamma^{\beta\nu}\dot{\gamma}_{\alpha\beta}\dot{\gamma}_{\mu\nu}
+(\frac{\dot{\gamma}}{\gamma})^{2}-4\gamma(R+2\Lambda_{c})=0\\
&\gamma^{\mu\rho}\dot{\gamma}_{\rho\nu}C^{\nu}_{\alpha\mu}-\gamma^{\mu\rho}
\dot{\gamma}_{\rho\alpha}C^{\nu}_{\mu\nu}=0\\
&(\gamma^{\alpha\rho}\dot{\gamma}_{\beta\rho})^{\cdot}-2\gamma(R^{\alpha}_{\beta}+2\Lambda_{c}\delta^{\alpha}_{\beta})=0
\end{align}
\end{subequations}
where tildes have been omitted. Finally some terminology may prove helpful; (\ref{quadraticconstraint})
is called \emph{Quadratic Constraint}, (\ref{linearconstraint}) are called \emph{Linear Constraints},
and (\ref{equationofmotion}) are simply the \emph{Equations of Motion}.
\section{The Abelian Model-Classical Consideration}
This model is characterized by the vanishing of all the structure
constants of the corresponding Lie Algebra. In particular, since
$C^{\alpha}_{\mu\alpha}=0$, it is also a Class A model. A choice
for the basis one-forms, in the spatial coordinate basis $(x,y)$,
is $\sigma^{\alpha}_{i}(x)=\delta^{\alpha}_{i}$ i.e. the
Kronecker's Delta. In matrix notation: \be
\sigma^{\alpha}_{i}(x)=\left(\begin{array}{cc}
  1 & 0 \\
  0 & 1
\end{array}\right)
\ee
The solution to the system (\ref{Aut}) is:
\begin{subequations}
\begin{align}
\Lambda^{\alpha}_{\beta} &\in GL(2,\mathbb{R})\\
P^{\alpha}(t)&=\left(x(t),y(t)\right)
\end{align}
\end{subequations}
A suitable basis, which spans the space of solutions to (\ref{generators}) is:
\be
\begin{array}{cccc}
  \lambda^{\alpha}_{\beta(1)}=\left(\begin{array}{cc}
    1 & 0 \\
    0 & -1
  \end{array}\right) & \lambda^{\alpha}_{\beta(2)}=\left(\begin{array}{cc}
    0 & 1 \\
    0 & 0
  \end{array}\right) & \lambda^{\alpha}_{\beta(3)}=\left(\begin{array}{cc}
    1 & 0 \\
    0 & 1
  \end{array}\right) & \lambda^{\alpha}_{\beta(4)}=\left(\begin{array}{cc}
    0 & 0 \\
    1 & 0
  \end{array}\right) \\
\end{array}
\ee It is noteworthy that  $P^{\alpha}(t)$ contains the entire
freedom carried by the Time Dependent A.I.D.'s, i.e. the two
arbitrary functions of time. Thus, since the matrix
$\Lambda^{\alpha}_{\beta}$ is constant, the only possible use of
this ``gauge'' freedom is, according to (\ref{gaugetransfor}),
to set the shift zero. Choosing the above mentioned time gauge, we
have:
\begin{subequations}
\begin{align}
&-\gamma^{\alpha\mu}\gamma^{\beta\nu}\dot{\gamma}_{\alpha\beta}\dot{\gamma}_{\mu\nu}
+(\frac{\dot{\gamma}}{\gamma})^{2}-8\gamma\Lambda_{c}=0\\
&\textrm{Linear Constraints are identically satisftied}\\
&(\gamma^{\alpha\rho}\dot{\gamma}_{\beta\rho})^{\cdot}-4\gamma\Lambda_{c}\delta^{\alpha}_{\beta}=0
\end{align}
\end{subequations}

\subsection{Abelian Model with vanishing Cosmological Constant}
The previous set of equations assumes the form:
\begin{subequations}\label{solutionAbNL}
\begin{align}
&(\vartheta^{\alpha}_{\alpha})^{2}-
\vartheta^{\alpha}_{\beta}\vartheta^{\beta}_{\alpha}=0\\
&\textrm{Linear Constraints are Identically Satisfied}\\
&\gamma^{\alpha\omega}\dot{\gamma}_{\omega\beta}-\vartheta^{\alpha}_{\beta}=0
\end{align}
\end{subequations}
where $\vartheta^{\alpha}_{\beta}$ is a constant matrix.
The integration of (\ref{solutionAbNL}) is straightforward. The
quadratic constraint, simply states that the determinant of the
matrix $\vartheta^{\alpha}_{\beta}$ vanishes; thus the initial
number of the independent components, is at most 3. But the
equations of motion, can be rewritten as: \be
\dot{\gamma}_{\alpha\beta}=\gamma_{\alpha\varrho}\vartheta^{\varrho}_{\beta}
\ee and a consistency requirement, emerges: \be
\dot{\gamma}_{\alpha\beta}=\dot{\gamma}_{\beta\alpha}\Rightarrow
\gamma_{\alpha\varrho}\vartheta^{\varrho}_{\beta}=
\gamma_{\beta\varrho}\vartheta^{\varrho}_{\alpha} \ee An
exhaustive consideration of all the cases concerning a matrix
$\vartheta^{\alpha}_{\beta}$, with vanishing determinant, plus the
consistency requirement, results in only the following two
distinct --at first sight-- solutions to the previous system: \be
\gamma_{\alpha\beta}=\left(\begin{array}{cc}
  1 & 0 \\
  0 & 1 \\
\end{array}\right)
\ee
with associated line element:
\be
(ds_{1})^{2}=-(dt)^{2}+(dx^{1})^{2}+(dx^{2})^{2}
\ee
and:
\be \gamma_{\alpha\beta}(\tau)=\left(\begin{array}{cc}
  1 & 0 \\
  0 & e^{\theta \tau} \\
\end{array}\right), ~\theta \in \mathbb{R}^{\ast}
\ee where $\mathbb{R}^{\ast}=\mathbb{R}-[0]$, with associated line
element: \be (ds_{2})^{2}=-e^{\theta
\tau}(d\tau)^{2}+(dy^{1})^{2}+e^{\theta \tau}(dy^{2})^{2}, ~\theta
\in \mathbb{R}^{\ast} \ee The transformation
$(t,x^{1},x^{2})\rightarrow(\tau,y^{1},y^{2})$: \be
(t,x^{1},x^{2})=\left(\frac{2}{\theta}\cosh(\frac{\theta
y^{2}}{2})e^{\theta\tau/2},
y^{1},\frac{2}{\theta}\sinh(\frac{\theta
y^{2}}{2})e^{\theta\tau/2}\right) \ee takes the line element
$(ds_{1})^{2}$ to $(ds_{2})^{2}$.

\subsection{Abelian Model with non vanishing Cosmological Constant}
Contraction of the equations of motion results in the integral of
motion: \be \label{integral of motion}
(\frac{\dot{\gamma}}{\gamma})^{2}-16\gamma\Lambda_{c}=2w=\textrm{const}
\ee Under the ''conformal'' change of depended variable
$\gamma_{\alpha\beta}=\overline{\gamma}_{\alpha\beta}\sqrt{\gamma}$
the equations of motion become: \be
\overline{\gamma}^{\alpha\omega}\dot{\overline{\gamma}}_{\omega\beta}-
\vartheta^{\alpha}_{\beta}=0 \ee and contraction of this set,
results in (since $\overline{\gamma}=1$)
$\vartheta^{\alpha}_{\alpha}=0$. Substitution of the quadratic
constraint, and usage of the integral of motion (\ref{integral of
motion}), yield: \be
w=\vartheta^{\alpha}_{\beta}\vartheta^{\beta}_{\alpha} \Rightarrow
w=2\vartheta^{1}_{1}+ \vartheta^{1}_{2}\vartheta^{2}_{1} \ee since
$\vartheta^{\alpha}_{\alpha}=0$.

Again, consideration of the equations of motion in terms of
$\overline{\gamma}_{\alpha\beta}$ for all the traceless matrices
$\vartheta^{\alpha}_{\beta}$, plus the consistency requirement
(i.e.
$\dot{\overline{\gamma}}_{\alpha\beta}=\dot{\overline{\gamma}}_{\beta\alpha}$)
produce the following two solutions: \be
\overline{\gamma}_{\alpha\beta}=\left(\begin{array}{cc}
  1 & 0 \\
  0 & 1 \\
\end{array}\right), ~\textrm{with} ~w=0
\ee and: \be
\overline{\gamma}_{\alpha\beta}(\tau)=\left(\begin{array}{cc}
  e^{\vartheta^{1}_{1}\tau} & 0 \\
  0 & e^{-\vartheta^{1}_{1}\tau} \\
\end{array}\right), ~\textrm{with} ~w=2(\vartheta^{1}_{1})^{2}>0,
~\vartheta^{1}_{1} \in \mathbb{R}^{\ast} \ee Since, for all the
cases, $w\geq 0$ the integral of motion (\ref{integral of motion})
dictates that $\Lambda_{c}>0$. Inversion of the conformal
transformation and integration of (\ref{integral of motion}) for
each of the two solutions, yield: \be
\gamma_{\alpha\beta}(t)=\frac{1}{2t\sqrt{\Lambda_{c}}}\left(\begin{array}{cc}
  1 & 0 \\
  0 & 1 \\
\end{array}\right)
\ee
with associated line element:
\be
(ds_{1})^{2}=-\frac{1}{4t^{2}\Lambda_{c}}(dt)^{2}+
\frac{1}{2t\sqrt{\Lambda_{c}}}((dx^{1})^{2}+(dx^{2})^{2})
\ee
and:
\be
\gamma_{\alpha\beta}(\tau)=\frac{\vartheta^{1}_{1}}{2\sqrt{\Lambda_{c}}}
\left(\begin{array}{cc}
  e^{\vartheta^{1}_{1}\tau} & 0 \\
  0 & e^{-\vartheta^{1}_{1}\tau} \\
\end{array}\right), ~\vartheta^{1}_{1} \in \mathbb{R}^{\ast}
\ee with associated line element: \be
(ds_{2})^{2}=-\frac{1}{4\Lambda_{c}\sinh^{2}(\tau)}(d\tau)^{2}+
\frac{e^{\tau}}{2\sqrt{\Lambda_{c}}\sinh(\tau)}(dy^{1})^{2}+
\frac{e^{-\tau}}{2\sqrt{\Lambda_{c}}\sinh(\tau)}(dy^{2})^{2} \ee
where $\vartheta^{1}_{1}$ , not being an essential constant
\cite{JMP}, has been absorbed.\\
It is obvious that the two line elements are connected by a time reparameterization of the
form $t\rightarrow \tau: ~t=\sinh(\tau)e^{-\tau}$.

\section{The Non Abelian Model-Classical Consideration}
This model is characterized by $C^{1}_{12}=1$; any other
choice can be cast into this. A choice for the basis
one-forms, in the spatial coordinate basis $(x^{1},x^{2})$ is: \be
\sigma^{\alpha}_{i}(x)=\left(\begin{array}{cc}
  e^{-2x^{2}} & 0 \\
  0 & 1
\end{array}\right)
\ee Now, the solution to the system (\ref{Aut}) is:
\begin{subequations}
\begin{align}
\Lambda^{\alpha}_{\beta}(t)&=\left(\begin{array}{cc}
  x(t) & y(t) \\
  0 & 1 \\
\end{array}\right)\\
P^{\alpha}(t)&=\left(\frac{\dot{y}(t)}{2}-\frac{y(t)\dot{x}(t)}{2x(t)},
-\frac{\dot{x}(t)}{2x(t)}\right)
\end{align}
\end{subequations}
This time, both the automorphic matrices
$\Lambda^{\alpha}_{\beta}(t)$ and the triplet $P^{\alpha}(t)$
carry the freedom contained in the time dependent A.I.D.'s. It is
wiser to exploit this freedom contained in
$\Lambda^{\alpha}_{\beta}(t)$ in order to simplify the initial
form of the scale factor matrix $\gamma_{\alpha\beta}(t)$ rather
than use the same freedom contained in $P^{\alpha}(t)$ to set the
shift vector
equal to zero.\\
Accordingly, an initial full scale factor matrix, can be brought
to the form: \be \gamma_{\alpha\beta}(t)=\left(\begin{array}{cc}
  \gamma_{11}(t) & 0 \\
  0 & \gamma_{11}(t)
\end{array}\right)
\ee Thus the initial set of ''dynamical'' variables, consists of a
scale factor matrix of the previous form, plus a shift vector
$(N^{1}(t),N^{2}(t))$. Insertion of this set into both
(\ref{linearconstraint}) and (\ref{curvature}) results in
respectively: \be N^{a}(t)=0 \ee and: \be
R^{\alpha}_{\beta}(t)=\frac{4}{\gamma_{11}(t)}\left(\begin{array}{cc}
  1 & 0 \\
  0 & 1
\end{array}\right)
\ee Thus, a judicious use of the gauge freedom can result in both
a very simple form of the scale factor matrix and the vanishing of
the shift. The equations of motion admit the integral of motion:
\be\label{secondIntegralofmotion}
(\frac{\dot{\gamma_{11}}}{\gamma_{11}})^{2}-(16\gamma_{11}+4\Lambda_{c}\gamma_{11}^{2})
=w=\textrm{const} \ee The quadratic constraint assumes the form:
\be
(\frac{\dot{\gamma_{11}}}{\gamma_{11}})^{2}-(16\gamma_{11}+4\Lambda_{c}\gamma_{11}^{2})
=0 \ee and thus not only sets the constant $w$ equal to zero, but
also demands $\Lambda_{c}>0$ --if this term exists. The
integration of (\ref{secondIntegralofmotion}) with $w=0$, is a
trivial matter:\\ When $\Lambda_{c}=0$ the result is \be
\gamma_{11}(t)=\frac{1}{4t^{2}} \ee  with corresponding line
element : \be (ds)^{2}=-\frac{1}{16t^{4}}(dt)^{2}+
\frac{e^{-4x^{2}}}{4t^{2}}(dx^{1})^{2}+\frac{1}{4t^{2}}(dx^{2})^{2}
~~~t \in \mathbb{R}^{\ast} \ee The transformation
$(t,x^{1},x^{2})\rightarrow(T,X,Y)$: \be
(T,X,Y)=\left(\frac{\cosh(2x^{2})}{4t}+2\frac{(x^{1})^{2}e^{-2x^{2}}}{4t},
\frac{\sinh(2x^{2})}{4t}+2\frac{(x^{1})^{2}e^{-2x^{2}}}{4t},
-2\frac{x^{1}e^{-2x^{2}}}{4t}\right) \ee
takes the standard Minkowski spacetime form to the line element above.\\
When $\Lambda_{c}\neq 0$ the result reads \be
\gamma_{11}(t)=\frac{4}{16t^{2}-\Lambda_{c}} \ee  the associated
line element being: \be
(ds)^{2}=-(\frac{4}{16t^{2}-\Lambda_{c}})^{2}(dt)^{2}+
\frac{4e^{-4x^{2}}}{16t^{2}-\Lambda_{c}}(dx^{1})^{2}+\frac{4}{16t^{2}
-\Lambda_{c}}(dx^{2})^{2}~~~t \in \mathbb{R}_{\star} \ee with
$\mathbb{R}_{\star}=\mathbb{R}-[-\sqrt{\Lambda_{c}}/4,
\sqrt{\Lambda_{c}}/4]$.
\newpage
\section{Quantum Description of the Models}
In trying to quantize gravity, one faces the problem of quantizing
a constrained system. The main steps one has to follow are:
\begin{itemize}
\item[(i)] define the basic operators $\widehat{g}_{ij}$ and
$\widehat{\pi}^{ij}$ and the canonical commutation relation
they satisfy.
\item[(ii)] define quantum operators $\widehat{H}_{m}$ whose classical counterparts
are the constraint functions $H_{m}$.
\item[(iii)] define the quantum states $\Psi[g]$ as the common null eigenvector of
$\widehat{H}_{m}$, i.e. those satisfying $\widehat{H}_{m}\Psi[g]=0$.
(As a consequence, one has to check that $\widehat{H}_{m}$, form a closed algebra
under the basic Canonical Commutation Relations (CCR).)
\item[(iv)] find the states and define the inner product in the space of these states.
\end{itemize}
Concerning point (iii) it is pertinent to clarify the meaning of
the imposition of the quantum constraints upon $\Psi[g]$. A
straightforward (modulo regularization prescriptions) but tedious
calculation shows that any functional which is not a scalar
functional of the curvature and/or higher derivative curvature
scalars  does not solve the linear operator constraints.
Therefore, the imposition of these conditions, ensures that the
wave functional will be a (scalar) functional of the 3-geometry
and not of the coordinate system on it. Then, the dynamical
evolution is provided by the quadratic operator constraint; the
consistency of the quantum algebra is, somehow, expected to
guarantee that the final wave functional will be independent of
the 4 dimensional coordinate system.

In the absence of a full solution to the problem, a partial
solution, generally known as quantum cosmology, has been employed.
This is an approximation to quantum gravity in which one freezes
out all but a finite number of degrees of freedom, and quantizes
the rest. In this way one is left with a much more manageable
problem that is essentially quantum mechanics with constraints. In
principle, the dynamical variables are the components of a 2x2
symmetric scale factor matrix $\gamma_{\alpha\beta}(t)$'s, the
lapse function $N(t)$ and the shift vector $N^{\alpha}(t)$. The
presence of the linear constraints --along with the conditional
symmetries of the corresponding Hamiltonian-- enable a reduction
of the initial configuration space to a lower dimensional one,
spanned by the curvature scalar characterizing the 2-geometry. The
ultimate justification of this reduction is the fact that --from
the point of view of the 2-geometry-- the omitted degrees of
freedom, are not physical but gauge \cite{CMP}. It is true that at
the classical level, the scale factor matrix, can be diagonalized
on mass-shell --through a constant matrix e.g. \cite{JMP} for the
3-dimensional analogous case-- while the shift can be set equal to
zero. However, if one intends to give weight to all states, one
has to start with the most general form which is described by the
3 scale factors $\gamma_{\alpha\beta}(t)$'s and the lapse function
$N(t)$. The absence of $H_{\alpha}$'s due to the vanishing of the
$C^{\alpha}_{\beta\gamma}$'s, implies that in principle all
$\gamma_{\alpha\beta}$'s are candidates as arguments for the wave
function which solves the quadratic constraint (Wheeler-DeWitt
equation). This is in contrast to what happens in the Non Abelian
case, where one combination of $\gamma_{\alpha\beta}$'s and
$C^{\alpha}_{\beta\gamma}$'s, parameterize the reduced
configuration space.

\subsection{Quantization of the Abelian Model}
In this section, we present a complete reduction of the initial
configuration space for the Abelian geometry --by extracting as
many gauge degrees of freedom, as possible. Two separate cases are
considered; when the cosmological constant is present and when is
not. In either case, a wave function which depends on one degree
of freedom is found, by imposing on it, the quantum versions of
all classical integrals of motion as additional conditions.

The Hamiltonian of the a Class A, spatially homogeneous
cosmological system is $H=N(t)H_{0}+N^{\alpha}(t)H_{\alpha}$,
where: \be \label{abelianhamiltonian}
H_{0}=\frac{1}{2\sqrt{\gamma}}L_{\alpha\beta\mu\nu}\pi^{\alpha\beta}\pi^{\mu\nu}+
\sqrt{\gamma}(R +2\Lambda) \ee is the quadratic constraint, with:
\be
\begin{array}{ll}\label{abeliansupermetricandR}
L_{\alpha\beta\mu\nu}=\gamma_{\alpha\mu}\gamma_{\beta\nu}
+\gamma_{\alpha\nu}\gamma_{\beta\mu}-2
\gamma_{\alpha\beta}\gamma_{\mu\nu} \\
R=C^{\beta}_{\lambda\mu}C^{\alpha}_{\theta\tau}\gamma_{\alpha\beta}
\gamma^{\theta\lambda}
\gamma^{\tau\mu}+2C^{\alpha}_{\beta\delta}C^{\delta}_{\nu\alpha}\gamma^{\beta\nu}+
4C^{\mu}_{\mu\nu}C^{\beta}_{\beta\lambda}\gamma^{\nu\lambda}
\end{array}
\ee $\gamma$ being the determinant of $\gamma_{\alpha\beta}$, and:
\be \label{abelianlinearconstraint}
H_{\alpha}=4C^{\mu}_{\alpha\rho}\gamma_{\beta\mu}\pi^{\beta\rho}
\ee are the linear constraints. For all Class A Types, the
canonical equations of motion, following from
(\ref{abelianhamiltonian}), are equivalent to Einstein's equations
derived from line element --see \cite{Sneddon} for the 3+1
dimensional analogous.

The quantities $H_{0}$, $H_{\alpha}$ are weakly vanishing
\cite{Dirac}, i.e. $H_{0}\approx 0$, $H_{\alpha}\approx 0$. For
the Class A ($C^{\alpha}_{\alpha\beta}=0$) Abelian Model, it can
be seen --using the basic Poisson Brackets Relations (PBR's)
$\{\gamma_{\alpha\beta},
\pi^{\mu\nu}\}=\delta^{\mu\nu}_{\alpha\beta}$-- that these
constraints are first class, obeying the following algebra: \be
\label{algebra}
\begin{array}{l}
  \{H_{0}, H_{\alpha}\}=0 \\
  \{H_{\alpha},
  H_{\beta}\}=-2C^{\gamma}_{\alpha\beta}H_{\gamma},
\end{array}
\ee which ensures their preservation in time, i.e.
$\dot{H}_{0}\approx 0$, $\dot{H}_{\alpha}\approx 0$, and
establishes the consistency of the action.

If we follow Dirac's general proposal \cite{Dirac} for quantizing
this action, we have to turn $H_{0}$, $H_{\alpha}$, into operators
annihilating the wave function $\Psi$.

In the Schr\"{o}dinger representation: \be
\begin{array}{l}
  \gamma_{\alpha\beta}\rightarrow
\widehat{\gamma}_{\alpha\beta}=\gamma_{\alpha\beta} \\
  \pi^{\alpha\beta}\rightarrow
\widehat{\pi}^{\alpha\beta}=-i\frac{\partial}{\partial\gamma_{\alpha\beta}},
\end{array}
\ee with the relevant operators, satisfying the basic Canonical
Commutation Relations\\ (CCR's) which correspond to the classical
ones: \be [\widehat{\gamma}_{\alpha\beta},
\widehat{\pi}^{\mu\nu}]=i\delta^{\mu\nu}_{\alpha\beta}=\frac{i}{2}
(\delta^{\mu}_{\alpha}\delta^{\nu}_{\beta}+\delta^{\mu}_{\beta}\delta^{\nu}_{\alpha}).
\ee

In the Abelian case, $C^{\alpha}_{\mu\nu}=0$, thus the only
operator which must annihilate the wave function, is
$\widehat{H}_{0}$; and the Wheeler-DeWitt equation
$\widehat{H}_{0}\Psi=0$, will produce a wave function, initially
residing on a 3-dimensional configuration space --spanned by
$\gamma_{\alpha\beta}$'s. If the linear constraints existed, a
first reduction of the initial configuration space, would be possible
\cite{Kuchar}. New variables, instead of the 3 scale
factors, would emerge --say $q^{i}$, with $i<3$. Then a new
``physical'' metric would be induced: \be \label{physicalmetric}
g^{ij}=L_{\alpha\beta\mu\nu}\frac{\partial q^{i} }{\partial
\gamma_{\alpha\beta}}\frac{\partial q^{j}}{\partial
\gamma_{\mu\nu}} \ee According to Kucha\v{r}'s and Hajicek's
\cite{Kuchar} prescription, the ``kinetic'' part of $H_{0}$ would
have to be realized as the conformal Laplacian (in order for the
equation to respect the conformal covariance of the classical
action), based on the physical metric (\ref{physicalmetric}). In
the presence of conditional symmetries, further reduction can take
place, a new physical metric would then be defined similarly, and
the above mentioned prescription, would have to be used after the
final reduction \cite{Kuchar2}.

The Abelian case, is an extreme example in which all the linear
constraints, vanish identically; thus no initial physical metric,
exists --another peculiarity reflecting the high spatial symmetry
of the model  under consideration. In compensation, a lot of
integrals of motion exist ant the problem of reduction, finds its
solution through the notion of \emph{``Conditional Symmetries''}.
These linear in momenta integrals of motion, if seen as vector
field on the configuration space spanned by
$\gamma_{\alpha\beta}$'s, induce --through their integral curves--
motions of the form $\widetilde{\gamma}_{\alpha\beta}=
\Lambda^{\mu}_{\alpha}\Lambda^{\nu}_{\beta}\gamma_{\mu\nu},
\Lambda \in GL(2,\mathbb{R})$ (see section 2 of \cite{CMP}) which not
only are identical to the action of spatial diffeomorphisms, but
also describe the action of the automorphism group --since
$GL(2,\mathbb{R})$ is the Aut(G) which corresponds to the Abelian models.

The generators of this automorphism group, are (in a collective
form and matrix notation) the following 4 --one for each
parameter: \be \label{lambdagenerators}
\lambda^{\alpha}_{(I)\beta}=\left(
\begin{array}{cc}
  \alpha  & \beta \\
  \delta & \varepsilon
\end{array}\right),~~~ I \in \{1,\ldots,4\}
\ee with the defining property: \be
C^{\alpha}_{\mu\nu}\lambda^{\kappa}_{\alpha}=
C^{\kappa}_{\mu\sigma}\lambda^{\sigma}_{\nu}+
C^{\kappa}_{\sigma\nu}\lambda^{\sigma}_{\mu}. \ee Exponentiating
all these matrices, one obtains the outer automorphism group of
the Abelian model, since there is not Inner Automorphism subgroup
(all structure constants vanish).

For full pure gravity, Kucha\v{r} \cite{Kuchar2} has shown that
there are no other first-class functions, homogeneous and linear
in the momenta, except the linear constraints --ditto in 2+1
analysis. If however, we impose extra symmetries, such quantities
may emerge --as it will be shown. We are therefore --according to
Dirac \cite{Dirac}-- justified to seek the generators of these
extra symmetries; their quantum-operator analogues will be imposed
as additional conditions on the wave function. The justification
for such an action, is obvious since these generators correspond
to spatial diffeomorphisms --which are the covariance of the
theory. Thus, these additional conditions are expected to lead us
to the final reduction, by revealing the true degrees of freedom.
Such quantities are, generally, called in the literature
\emph{``Conditional Symmetries''} \cite{Kuchar2}.

From matrices (\ref{lambdagenerators}), we can construct the
linear --in momenta-- quantities: \be \label{epsilons}
E_{(I)}=\lambda^{\alpha}_{(I)\beta}\gamma_{\alpha\rho}\pi^{\rho\beta},~~~
I \in \{1,\ldots,4\} \ee In order to write analytically these
quantities, the following base is chosen: \be \label{basis}
\begin{array}{cccc}
 \lambda_{1}= \left(\begin{array}{cc}
  1 & 0 \\
  0 & -1
\end{array}\right), & \lambda_{2}=\left(\begin{array}{cc}
  0 & 1 \\
  0 & 0
\end{array}\right), & \lambda_{3}=\left(\begin{array}{cc}
  0 & 0 \\
  1 & 0
\end{array}\right), & \lambda_{4}=\left(\begin{array}{cc}
  1 & 0 \\
  0 & 1
\end{array}\right)
\end{array}
\ee

It is straightforward to calculate the Poisson Brackets between
$E_{(I)}$ and $H_{0}$: \be \label{commutatorEH} \{E_{(I)},
H_{0}\}=-2\Lambda N\sqrt{\gamma}\lambda^{a}_{(I)a} \ee But, it
holds that: \be \dot{E}_{(I)}=\{E_{(I)},H_{0}\}=-2\Lambda
N\sqrt{\gamma}\lambda^{a}_{(I)a}\ee --the last equality emerging
by virtue of (\ref{commutatorEH}). Thus: \be
\label{integralsofmotionforE} \dot{E}_{(I)}=\{E_{(I)},H_{0}\}=0
\Rightarrow E_{(I)}=K_{(I)}=\textrm{constants},~~~ I \in \{1,2,3\}
\ee We therefore conclude that, when the cosmological constants is
non vanishing, only the first three $E_{(I)}$, are first-class and
thus integrals of motion. If $\Lambda$ vanishes, all the four
quantities, are first-class. Out of the three quantities
$E_{(I)}$, only two are (functionally) independent , if we allow
for the coefficients of the linear combination to be functions of
the $\gamma_{\alpha\beta}$'s; if the coefficients are only allowed
to be numbers , all three are (linearly) independent.

The algebra of $E_{(I)}$ can be easily seen to be: \be
\label{algebraofepsilons}
\{E_{(I)},E_{(J)}\}=-\frac{1}{2}C^{M}_{IJ}E_{(M)},~~~I,J,M \in
\{1,\ldots,4\} \ee where: \be \label{algebraoflambda}
[\lambda_{(I)},\lambda_{(J)}]=C^{M}_{IJ}\lambda_{(M)},~~~I,J,M \in
\{1,\ldots,4\} \ee the square brackets denoting matrix
commutation.\\
The non vanishing structure constants of the algebra
(\ref{algebraoflambda}), are found to be: \be
\label{liealgebraoflambda}
\begin{array}{ccc}
  C^{2}_{12}=2 & C^{3}_{13}=-2 & C^{1}_{23}=1
\end{array}
\ee

At this point, in order to achieve the desired reduction, we
propose that the quantities $E_{(I)}$ --with $I \in
\{1,\ldots,3\}$-- must be promoted to operational conditions
acting on the requested wave function $\Psi$ --since they are
first class quantities and thus integrals of motion (see
(\ref{integralsofmotionforE})). In the Schr\"{o}dinger
representation: \be \label{epsilonuponPsi}
\widehat{E}_{(I)}\Psi=-i\lambda^{\tau}_{(I)\alpha}\gamma_{\tau\beta}\frac{\partial
\Psi}{\partial \gamma_{\alpha\beta}}=K_{(I)}\Psi,~~~I \in
\{1,\ldots,3\} \ee In general, systems of equations of this type,
must satisfy consistency conditions decreed by the Frobenious
Theorem: \be
\begin{array}{ccc}
  \widehat{E}_{(J)}\Psi=K_{(J)}\Psi & \Rightarrow &
  \widehat{E}_{(I)}\widehat{E}_{(I)}\Psi=K_{(I)}K_{(J)}\Psi\\
  \widehat{E}_{(I)}\Psi=K_{(I)}\Psi & \Rightarrow &
  \widehat{E}_{(J)}\widehat{E}_{(I)}\Psi=K_{(J)}K_{(I)}\Psi
\end{array}
\ee Subtraction of these two and usage of
(\ref{algebraofepsilons}), results in: \be \label{selectionrule}
K^{M}_{IJ}\widehat{E}_{(M)}\Psi=0 \Rightarrow
C^{M}_{IJ}K_{(M)}=0\ee This relation constitutes a selection rule
for the numerical values of the integrals of motion. In view of
the Lie Algebra (\ref{liealgebraoflambda}), selection rule
(\ref{selectionrule}) sets $K_{1}=K_{2}=K_{3}=0$. This fact
restores the action of the diffeomorphisms as covariances of the
quantum theory, in the sense that now, we have conditions of the
form $\widehat{E}_{(I)}\Psi=0$. On the contrary, if we also had $E_{(4)}$
(as is the case $\Lambda=0$) then $K_{4}$ would remain arbitrary.
With this outcome, and using the method of characteristics
\cite{Carabedian}, the system of the two functionally independent
P.D.E.'s (\ref{epsilonuponPsi}), can be integrated. The result is:
 \be \label{wavefunction}
\Psi=\Psi(\gamma) \ee i.e. an arbitrary function of $\gamma$ --the
determinant of the scale factor matrix. Note that the solution
(\ref{wavefunction}) would have not changed if an other base for
the generators had been chosen. However, such a choice would have
had affected the form of the system (\ref{epsilonuponPsi}).

The next step, is to construct the Wheeler-DeWitt equation which
is to be solved by the wave function (\ref{wavefunction}). The
degree of freedom, is 1; the $q=\gamma$. According to Kucha\v{r}'s
proposal \cite{Kuchar}, upon quantization, the kinetic part of
Hamiltonian is to be realized as the conformal Beltrami operator
-- based on the induced physical metric --according to
(\ref{physicalmetric}), with $q=\gamma$: \be g^{11}=
L_{\alpha\beta\mu\nu}\frac{\partial \gamma }{\partial
\gamma_{\alpha\beta}}\frac{\partial \gamma}{\partial
\gamma_{\mu\nu}}=L_{\alpha\beta\mu\nu}\gamma^{2}\gamma^{\alpha\beta}\gamma^{\mu\nu}
\stackrel{\textrm{first of~}
(\ref{abeliansupermetricandR})}{=}-4\gamma^{2}\ee In the
Schr\"{o}dinger representation: \be
\frac{1}{2}L_{\alpha\beta\mu\nu}\pi^{\alpha\beta}\pi^{\mu\nu}
\rightarrow -\frac{1}{2}\Box^{2}_{c} \ee where: \be \label{box}
\Box^{2}_{c}=\Box^{2}=\frac{1}{\sqrt{g_{11}}}~\partial_{\gamma}
\{\sqrt{g_{11}}~g^{11}~\partial_{\gamma}\} \ee is the
1--dimensional Laplacian based on $g_{11}$ ($g^{11}g_{11}=1$).
Note that in 1--dimension the conformal group is totally contained
in the G.C.T. group, in the sense that any conformal
transformation of the metric can not produce any change in the
--trivial-- geometry and is thus reachable by some G.C.T.
Therefore, no extra term in needed in (\ref{box}), as it can also
formally be seen by taking the limit $n=1,~R=0$ in the general
definition: \be
\Box^{2}_{c}\equiv\Box^{2}+\frac{(d-2)}{4(d-1)}R=\Box^{2} \ee
Thus: \be H_{0}\rightarrow
\widehat{H}_{0}=2\gamma\partial_{\gamma}(\gamma\partial_{\gamma})+2\Lambda
\gamma\ee So, the Wheeler-DeWitt equation --by virtue of
(\ref{wavefunction})--, reads: \be
\widehat{H}_{0}\Psi=\gamma^{2}\Psi''+\gamma\Psi'+\gamma\Lambda\Psi=0
\ee The general solution to this equation, is: \be
\Psi(\gamma)=c_{1}J_{0}(2\sqrt{\gamma\Lambda})
+c_{2}Y_{0}(2\sqrt{\gamma\Lambda}) \ee where $J_{n}$ and $Y_{n}$,
are the Bessel Functions of the first and second kind respectively
--both of zero order-- and $c_{1},c_{2}$, arbitrary constants.
Some comments on this wave function. Indeed, at first sight, the
fact that $\Psi$ depends only on one argument and particularly on
$\gamma$, seems to point to some undesirable degeneracy regarding
anisotropy; classically $\gamma$ can be gauged to $e^{t}$ and thus
it seems as though the anisotropy parameter does not enter $\Psi$
at all. If, however, we reflect thoroughly, we will realize that
this objection rests strongly on a --not generally accepted--
mingling of the classical notion of anisotropy and the
interpretation of the wave function. Indeed if we adopt the
interpretation that the wave function $\Psi$ (along with a
suitable measure), is to give weight to all configurations
parameterized by $\gamma_{\alpha\beta}$, then the anisotropic
configuration will, in general, acquire different probabilities.
The degeneracy occurs only between two different anisotropic
configurations with the same determinant $\gamma$. In compensation
the scheme proposed here, avoids the gauge degrees of freedom as
much as possible. The final probabilistic interpretation must
await the selection of a proper measure.

If the cosmological constant $\Lambda$ is zero, some changes will
take place. The first concerns the obvious alteration to the
potential in the Hamiltonian; it vanishes. This consequently,
causes an alteration to the Poisson Bracket (\ref{commutatorEH}),
which takes the form: \be \label{alteredcommutatorEH}
\{E_{(I)},H_{0}\}=0,~~~I \in \{1,\ldots,4\} \ee while,
(\ref{algebraofepsilons}) still holds. Thus in the present case,
there are four integrals of motion --instead of three. Also, the
P.D.E. system (\ref{epsilonuponPsi}) consists of four members
(instead of three), but now out of the four quantities $E_{(I)}$,
only three are functionally independent; the previous two, plus
$E_{(4)}$. Again, using the method of characteristics
\cite{Carabedian}, the system of the three functionally
independent P.D.E.'s (\ref{epsilonuponPsi}), can be integrated.
The result is: \be \label{secondwavefunction}
\Psi=c_{1}\gamma^{iK_{4}/2} \ee where $\gamma$ is the determinant
of the scale factor and $K_{4}$, the remaining constant
--according to selection rule (\ref{selectionrule}).

The fact that this wave function does not depend on any
combination of $\gamma_{\alpha\beta}$'s in an arbitrary manner
(i.e. $\Psi$ is not an arbitrary function of
$\gamma_{\alpha\beta}$'s), might be taken as an indication that no
reduced Wheeler-DeWitt equation can be written. On the other hand,
this wave function does contain an arbitrary  constant which, at
the classical level is not essential ( the model is 2+1 Minkowski
spacetime). Thus, it would not be harmful if the value of this
constant were  to be fixed by the quantum dynamics. These thoughts
lead to the following compromise; the initial configuration space,
should be the mini-superspace i.e. we should write the
Wheeler-DeWitt equation, based on the supermetric
$L^{\alpha\beta\mu\nu}$.

In the Schr\"{o}dinger representation: \be
\frac{1}{2}L_{\alpha\beta\mu\nu}\pi^{\alpha\beta}\pi^{\mu\nu}
\rightarrow -\frac{1}{2}\Box^{2}_{c} \ee  Thus using
(\ref{superricciscalar}), (\ref{superlinearterm}),
(\ref{superfinalBeltrami}) for $n=2$ and $\mathcal{D}=3$, one may
find respectively --see appendix: \be R=2 \ee \be
L_{\alpha\beta\mu\nu}\Gamma^{\alpha\beta\mu\nu}_{\kappa\lambda}=
-\gamma_{\kappa\lambda} \ee and: \be \Box^{2}_{c}=
L_{\alpha\beta\mu\nu}\frac{\partial}{\partial
\gamma_{\alpha\beta}\gamma_{\mu\nu}}+\gamma_{\kappa\lambda}\frac{\partial}{\partial
\gamma_{\kappa\lambda}}+\frac{1}{4} \ee Then Kucha\v{r}'s proposal
for the Hamiltonian reads: \be H_{0}\rightarrow
\widehat{H}_{0}=-\frac{1}{2}\left(L_{\alpha\beta\mu\nu}\frac{\partial}{\partial
\gamma_{\alpha\beta}\gamma_{\mu\nu}}+\gamma_{\kappa\lambda}\frac{\partial}{\partial
\gamma_{\kappa\lambda}}+\frac{1}{4}\right) \ee

Substitution of the wave function (\ref{secondwavefunction}) in
the Wheeler-DeWitt equation $\widehat{H}_{0}\Psi=0$, with
$\widehat{H}_{0}$ given by the previous relation, determines the
constant $K_{4}$. The outcome is: \be
\Psi=c_{1}\gamma^{1/4}+c_{2}\gamma^{-1/4} \ee The constants
$c_{1}, c_{2}$, remain arbitrary and may be fixed after the
selection of a proper measure via normalizability requirements.

\newpage
\subsection{Quantization of the Non Abelian Model}
In the present section, the quantization of the Non Abelian model,
is exhibited. This model is a Class B since $C^{\mu}_{\mu\nu}\neq
0$. Class B Cosmological Models, have a peculiarity; if the
simplifying hypothesis of homogeneity, is inserted in the
Einstein-Hilbert action, the reduced action obtained, will result
in equations of motion which are, in general, not equivalent to
the equations one gets by the imposition of the same hypothesis,
directly on the full Einstein's Field Equations. This situation
does not occur for the case of Class A models. Suppose that one
adopts the canonical analysis in the framework of the Hamiltonian
description. Then, the problem of the existence of a ``valid''
Hamiltonian (i.e. of a Hamiltonian which produces equations of
motion equivalent to the corresponding Einstein equations),
arises. A great many of works have dealt with the problem
(\cite{ValidHamiltonians} and the references therein). The
conclusion was that for Class B spacetimes, with a general scale
factor matrix $\gamma_{\alpha\beta}(t)$, a valid Hamiltonian is
not known --a serious drawback since one major aim of the
Hamiltonian approach, is the quantization of the system under
discussion.\\
Though it is extremely difficult to attack this problem, partial
solutions have been given in \cite{ValidHamiltonians}. Indeed, in
that work, a Hamiltonian constructed out of the scale factor
matrix $\gamma_{\alpha\beta}(t)$ and the structure constants
$C^{\alpha}_{\mu\nu}$, which resembles in form the Hamiltonian for
the Class A models, is constructed and sufficient conditions on
the various parts of that Hamiltonian are given, in order for it
to be valid. The set of these conditions is large, the
conditions themselves a little complicated and auxiliary
quantities enter the scheme. But, in 2+1 analysis, not all are
needed; the equations of motion, are nothing but the derivatives of the linear
constraints --a fact which simplifies the system of the conditions to be satisfied and
the procedure of identifications.

Thus, following the entire procedure described in
\cite{ValidHamiltonians}, the following results are obtained:

The valid Hamiltonian for the Non Abelian Model, is: \be
H=N(t)\Big(\frac{1}{2}\Theta_{\alpha\beta\mu\nu}\pi^{\alpha\beta}\pi^{\mu\nu}+V\Big)+4
N^{\rho}(t)C^{\kappa}_{\rho\alpha}\gamma_{\kappa\beta}(t)\pi^{\alpha\beta}
\ee where:
\begin{subequations}
\begin{align}
\Theta_{\alpha\beta\mu\nu}&=\frac{1}{6q}(\Sigma_{\alpha\beta}\gamma_{\mu\nu}+
\Sigma_{\mu\nu}\gamma_{\alpha\beta})-\frac{1}{6}(\gamma_{\alpha\mu}\gamma_{\beta\nu}+
\gamma_{\alpha\nu}\gamma_{\beta\mu})+\gamma_{\alpha\beta}\gamma_{\mu\nu}\\
\Sigma_{\alpha\beta}&=C^{\tau}_{\tau\xi}(C^{\omega}_{\phi\alpha}\gamma_{\omega\beta}
+C^{\omega}_{\phi\beta}\gamma_{\omega\alpha})\gamma^{\xi\phi}\\
q&=C^{\alpha}_{\beta\mu}C^{\beta}_{\alpha\nu}\gamma^{\mu\nu}\\
V&=-12R-3\Lambda
\end{align}
\end{subequations}
The quantity q is scalar under the action of the automorphism
group, corresponds to the unique curvature scalar of the spatial
surfaces of simultaneity $\Sigma_{t}$ ($q=\frac{R}{8}$) and thus
exhibits the only true degree of freedom --as far as the
2-geometry is concerned.

The corresponding scalar (under the action of the automorphism
group) Lagrangian is: \be
L=\frac{1}{2N(t)}\Big(\Theta^{\alpha\beta\mu\nu}
\mathcal{K}_{\alpha\beta}\mathcal{K}_{\mu\nu}\Big)-N(t)V \ee where
$\Theta^{\alpha\beta\mu\nu}$ is the inverse of
$\Theta_{\alpha\beta\mu\nu}$: \be
\Theta^{\alpha\beta\mu\nu}\Theta_{\mu\nu\kappa\lambda}=
\delta^{\alpha\beta}_{\kappa\lambda} \ee given by:
\begin{subequations}
\begin{align}
\Theta^{\alpha\beta\mu\nu}&=\Delta^{\alpha\beta\mu\nu}+
c_{1}(S^{\alpha\beta}G^{\mu\nu}+S^{\mu\nu}G^{\alpha\beta})
+c_{2}(S^{\alpha\beta}S^{\mu\nu})+c_{3}(G^{\alpha\beta}G^{\mu\nu})\\
\omega&=\frac{1}{6q}\\
c_{1}&=-\frac{\omega(1+\Gamma\omega)}{1+2\Gamma\omega+\Gamma^{2}\omega^{2}
-GS\omega^{2}}\\
c_{2}&=\frac{G\omega^{2}}{1+2\Gamma\omega+\Gamma^{2}\omega^{2}
-GS\omega^{2}}\\
c_{3}&=-\frac{S\omega^{2}}{GS\omega^{2}-(1+\Gamma\omega)^{2}}\\
\Delta^{\alpha\beta\mu\nu}&=-\frac{3}{2}(\gamma^{\alpha\mu}\gamma^{\beta\nu}+
\gamma^{\alpha\nu}\gamma^{\beta\mu})+\frac{9}{5}\gamma^{\alpha\beta}\gamma^{\mu\nu}\\
S^{\alpha\beta}&=\Delta^{\alpha\beta\mu\nu}\Sigma_{\mu\nu}\\
G^{\alpha\beta}&=\Delta^{\alpha\beta\mu\nu}\gamma_{\mu\nu}\\
S&=\Delta^{\alpha\beta\mu\nu}\Sigma_{\alpha\beta}\Sigma_{\mu\nu}\\
G&=\Delta^{\alpha\beta\mu\nu}\gamma_{\alpha\beta}\gamma_{\mu\nu}\\
\Gamma&=\Delta^{\alpha\beta\mu\nu}\gamma_{\alpha\beta}\Sigma_{\mu\nu}
\end{align}
\end{subequations}
and: \be
\mathcal{K}_{\alpha\beta}(t)=\dot{\gamma}_{\alpha\beta}(t)-
2N^{\phi}(t)(C^{\rho}_{\phi\alpha}
\gamma_{\rho\beta}+C^{\rho}_{\phi\beta} \gamma_{\rho\alpha}) \ee
while: \be
\pi^{\alpha\beta}=\Theta^{\alpha\beta\mu\nu}\mathcal{K}_{\mu\nu}\ee
by inversion of the equation: \be
\dot{\gamma}_{\alpha\beta}=\frac{\partial H}{\partial
\pi^{\alpha\beta}} \ee

Again upon quantization, following Kucha\v{r}'s proposal, in the
Schr\"{o}dinger representation: \be
\frac{1}{2}\Theta_{\alpha\beta\mu\nu}\pi^{\alpha\beta}\pi^{\mu\nu}
\rightarrow -\frac{1}{2}\Box^{2}_{c} \ee where: \be
\label{boxnonabelian}
\Box^{2}_{c}=\Box^{2}=\frac{1}{\sqrt{g_{11}}}~\partial_{\gamma}
\{\sqrt{g_{11}}~g^{11}~\partial_{\gamma}\} \ee is the
1--dimensional Laplacian based on the ``physical metric''
$g_{11}$: \be g^{11}=\Theta_{\alpha\beta\mu\nu}\frac{\partial
q}{\partial \gamma_{\alpha\beta}}\frac{\partial q}{\partial
\gamma_{\mu\nu}}=\frac{2}{3}q^{2} \ee with $g^{11}g_{11}=1$
--similarly to the Abelian case. It has been mentioned that in
1--dimension the conformal group is totally contained in the
G.C.T. group, in the sense that any conformal transformation of
the metric can not produce any change in the --trivial-- geometry
and is thus reachable by some G.C.T. Therefore, no extra term in
needed in (\ref{boxnonabelian}), as it can also formally be seen
by taking the limit $n=1,~R=0$ in the general definition: \be
\Box^{2}_{c}\equiv\Box^{2}+\frac{(d-2)}{4(d-1)}R=\Box^{2} \ee
Thus: \be H_{0}\rightarrow
\widehat{H}_{0}=-\frac{1}{3}q\partial_{q}(q\partial_{q})-3(4q+\Lambda)\ee
So, the Wheeler-DeWitt equation now, reads: \be
\widehat{H}_{0}\Psi=-\frac{q^{2}}{3}\Psi''(q)-\frac{q}{3}\Psi'(q)
-12q\Psi(q)-3\Lambda\Psi(q)=0 \ee The general solution to this
equation, for $\Lambda\neq 0$,is:
\be
\Psi(\gamma)=c_{1}J_{6i\sqrt{\Lambda}}(12\sqrt{q})
+c_{2}J_{-6i\sqrt{\Lambda}}(12\sqrt{q}) \ee where $J_{n}$, is the
Bessel Function of the first kind and $c_{1},c_{2}$, arbitrary
constants.

If $\Lambda$ vanishes, the solution is: \be
\Psi(\gamma)=c_{3}J_{0}(12\sqrt{q}) +c_{4}Y_{0}(12\sqrt{q}) \ee
where $Y_{n}$ is the Bessel Function of the second kind, and
$c_{3},c_{4}$, arbitrary constants.

\subsection{The Equivalence of the State Spaces}
In our examples the classical Geometries are either Minkowski
spaces or Spaces of Constant curvature, which in three dimensions
implies maximal symmetry. Thus, the 3 curvature invariants suffice
to characterize the space. We therefore take the following set of
invariant relations:
\begin{subequations}
\begin{align}
Q_{1}&=R^{A}_{A}\\
Q_{2}&=R^{A}_{B}R^{B}_{A}-\frac{R^{2}}{3}\\
Q_{3}&=R^{A}_{B}R^{B}_{\Gamma}R^{\Gamma}_{A}-\frac{R^{3}}{9}
\end{align}
\end{subequations}

Now, in Gauss normal coordinates, a three dimensional spatially
homogeneous  metric takes the form: \be
g_{AB}=\left(\begin{array}{ccc}
  -1 & 0 & 0 \\
  0 & g_{11} & g_{12}\\
  0 & g_{12} & g_{22}\\
\end{array}\right)
\ee where the spatial part $g_{ij}$ is: \be
g_{ij}=\sigma^{\alpha}_{i}\sigma^{\beta}_{j}\gamma_{\alpha\beta}
\ee  For the scale factor matrix $\gamma_{\alpha\beta}$ we assume
three arbitrary functions of time, i.e. we depart from the
classical solutions yet keeping ourselves within the class of
spatially homogeneous three-geometries.If we now use the one-forms
appropriate for the Abelian and non-Abelian symmetry group, we
get, as expected from the theorem of section 2, the following
results concerning the pair designated by the vanishing of the
cosmological constant(containing the two cosmological
parameterizations of 3d Minkowski space):
\paragraph{Abelian, $\Lambda=0$}
\begin{subequations}
\begin{align}
Q_{1}&=-2G^{0}_{0}\\
Q_{2}&=\frac{2}{3}(G^{0}_{0})^{2}\\
Q_{3}&=-\frac{10}{9}(G^{0}_{0})^{3}
\end{align}
\end{subequations}

\paragraph{Non Abelian, $\Lambda=0$}
\begin{subequations}
\begin{align}
Q_{1}&=-2G^{0}_{0}\\
Q_{2}&=\frac{2}{3}(G^{0}_{0})^{2}-2\gamma^{\alpha\beta}G^{0}_{\alpha}G^{0}_{\beta}\\
Q_{3}&=-\frac{10}{9}(G^{0}_{0})^{3}+3\gamma^{\alpha\beta}G^{0}_{\alpha}G^{0}_{\beta}G^{0}_{0}
\end{align}
\end{subequations}

\newpage
while, when a cosmological constant exists, the corresponding pair
(containing the two cosmological parameterizations of 3d Maximally
Symmetric space) gives the following set of relations:
\paragraph{Abelian, $\Lambda\neq 0$}
\begin{subequations}
\begin{align}
Q_{1}&=-2(G^{0}_{0}+\Lambda)\\
Q_{2}&=\frac{2}{3}(G^{0}_{0}+\Lambda)^{2}\\
Q_{3}&=-\frac{10}{9}(G^{0}_{0}+\Lambda)^{3}+4\Lambda(G^{0}_{0}+\Lambda)^{2}
\end{align}
\end{subequations}

\paragraph{Non Abelian, $\Lambda\neq 0$}
\begin{subequations}
\begin{align}
Q_{1}&=-2(G^{0}_{0}+\Lambda)\\
Q_{2}&=\frac{2}{3}(G^{0}_{0}+\Lambda)^{2}-2\gamma^{\alpha\beta}G^{0}_{\alpha}G^{0}_{\beta}\\
Q_{3}&=-\frac{10}{9}(G^{0}_{0}+\Lambda)^{3}+4\Lambda(G^{0}_{0}+\Lambda)^{2}+3(G^{0}_{0}+\Lambda)
\gamma^{\alpha\beta}G^{0}_{\alpha}G^{0}_{\beta}-12\Lambda\gamma^{\alpha\beta}G^{0}_{\alpha}G^{0}_{\beta}
\end{align}
\end{subequations}

Now, it is a fact that $G^{0}_{0}$ and $G^{0}_{\alpha}$ become,
upon transition to a Hamiltonian formulation, linear combinations
of $\mathcal{H}_{0}$ and $\mathcal{H}_{\alpha}$. This, in turn,
permits us to conclude that, upon canonically quantizing the pairs
in the two cases above mentioned, there will always exist an
entire host of factor orderings for each of the quantum analogues
of $Q_{1},Q_{2},Q_{3}$ such that their eigenvalues become $0$ or
$6\Lambda$ (for $Q_{1}$). This establishes the equivalence of the
corresponding quantum states under non-trivial space-time
coordinate transformations. Note that, when an arbitrary gauge is
used, the previous relations simply become more complicated; the
qualitative result that $Q_{1},Q_{2},Q_{3}$ are functions of the
constraints, remains of course valid.

\section{Discussion}
We have discussed the issue of space-time covariance of Canonical
Quantization of General Relativity. To many this is known as the
problem of time. At the classical level, the well established
canonical analysis of the Einstein-Hilbert action leaves no room
for doubts: the formulation is explicitly  space-time generally
covariant. Upon canonically quantizing the problem seems to
reappear as all the ingredients of the theory, i.e. the quantum
constraints and consequently  the quantum states concern the
three-geometry and not the space-time in which it is embedded. An
answer to the problem presupposes a commitment about what will
constitute the set of observables. Motivated by the very essence
of the notion of a geometry, we adopt the point of view that this
set must be identified by all invariant relations  between the
various curvature or higher derivative curvature scalars of a
given geometry.Each such  relation must be turned into an entity
living on the phase space by eliminating the time derivatives of
the extrinsic curvature (through use of the spatial equations of
motion, if necessary). The appeal to a well known theorem of
Constrained Dynamics permits  us to conclude that all these
entities are homogeneous polynomials of the Quadratic and the
Linear constraints. Therefore, when we wish to turn each and every
such entity into operator ( Quantum Observable), there will be
many factor orderings (namely all these that keep at least one
constraint to the far right) which will enable this operator to
annihilate the states defined as  the common null eigenstates of
the Quantum Constraints. Consequently, the use of  many different
slices as bases for canonically quantizing one and the same
space-time can have no effect on the so defined quantum
observables. Space-time covariance is thus observed at the quantum
level. The above considerations are not meant to imply that we
claim we have constructed a Dirac Consistent quantum theory of
General Relativity. They rather point to the claim that, if a
consistent imposition of the quantum constraints is achieved, we
expect to encounter no additional problems concerning space-time
covariance of the ensuing theory. In that sense, our result for
full pure gravity is formal. It is seen to be explicitly realized
in the case of 2+1 cosmology considered.
\newpage
\vspace{1.5cm} \noindent \textbf{Acknowledgements}\\
The authors acknowledge a vivid discussion with Prof. Rafael
Sorkin, whose criticism helped them shape the notion of Quantum
Equivalence. They also thank for discussion Dr. E. Korfiatis and
Prof. J. Zanelli. \\This work has been supported by EPEAEK II in
the framework of "PYTHAGORAS II - SUPPORT OF RESEARCH GROUPS IN
UNIVERSITIES".
\newpage
\appendix
\section{Appendix}
In this appendix, we give some useful formulae, concerning the
mini-superspace. \vspace{0.5cm}

Using the results of canonical analysis in a $(n+1)$--dimensional
manifold, endowed with the line element, one arrives at the notion
of mini-superspace spanned by $\gamma_{\alpha\beta}$'s
(co-ordinates), and having as ``covariant'' metric the following:
\be \label{supercovariantmetric}
L^{\alpha\beta\mu\nu}=\frac{1}{4}(\gamma^{\alpha\mu}\gamma^{\beta\nu}
+\gamma^{\alpha\nu}\gamma^{\beta\mu}
-2\gamma^{\alpha\beta}\gamma^{\mu\nu}) \ee while the
``contravariant metric'', is defined as: \be
\label{supercontravariantmetric}
L_{\alpha\beta\mu\nu}=(\gamma_{\alpha\mu}\gamma_{\beta\nu}
+\gamma_{\alpha\nu}\gamma_{\beta\mu}
-\frac{2}{n-1}\gamma_{\alpha\beta}\gamma_{\mu\nu}) \ee in the
sense that: \be \label{superdelta}
L^{\alpha\beta\kappa\lambda}L_{\kappa\lambda\mu\nu}=\delta^{\alpha\beta}_{\mu\nu}
\equiv \frac{1}{2}(\delta^{\alpha}_{\mu}\delta^{\beta}_{\nu}+
\delta^{\alpha}_{\nu}\delta^{\beta} _{\mu}) \ee

The ``Christoffel'' symbols are defined as: \be
\label{superchistoffel}
\Gamma^{\alpha\beta\mu\nu}_{\kappa\lambda}=\frac{1}{2}L_{\kappa\lambda\rho\sigma}
\{L^{\rho\sigma\mu\nu,\alpha\beta}+L^{\alpha\beta\rho\sigma,\mu\nu}
-L^{\alpha\beta\mu\nu,\rho\sigma}\}\ee where: \be
L^{\alpha\beta\mu\nu,\rho\sigma}\equiv \frac{\partial
L^{\alpha\beta\mu\nu}}{\partial \gamma_{\rho\sigma}} \ee Combined
usage of (\ref{supercovariantmetric}),
(\ref{supercontravariantmetric}), (\ref{superdelta}) and
(\ref{superchistoffel}), gives: \be
\Gamma^{\alpha\beta\mu\nu}_{\kappa\lambda}=-\frac{1}{4}(
\gamma^{\alpha\mu}\delta^{\beta\nu}_{\kappa\lambda}+
\gamma^{\alpha\nu}\delta^{\beta\mu}_{\kappa\lambda}+
\gamma^{\beta\mu}\delta^{\alpha\nu}_{\kappa\lambda}+
\gamma^{\beta\nu}\delta^{\alpha\mu}_{\kappa\lambda}) \ee

In the same spirit, ``Riemann'' tensor is defined as follows: \be
R^{\alpha\beta\rho\sigma\mu\nu}_{\kappa\lambda}=
\Gamma^{\alpha\beta\rho\sigma,\mu\nu}_{\kappa\lambda}-
\Gamma^{\alpha\beta\mu\nu,\rho\sigma}_{\kappa\lambda}+
\Gamma^{\alpha\beta\rho\sigma}_{\omega\xi}
\Gamma^{\omega\xi\mu\nu}_{\kappa\lambda}-
\Gamma^{\alpha\beta\mu\nu}_{\omega\xi}
\Gamma^{\omega\xi\rho\sigma}_{\kappa\lambda} \ee where: \be
\Gamma^{\alpha\beta\mu\nu,\rho\sigma}_{\kappa\lambda}\equiv
\frac{\partial
\Gamma^{\alpha\beta\mu\nu}_{\kappa\lambda}}{\partial
\gamma_{\rho\sigma}} \ee Contraction of $(\rho,\sigma)$ with
$(\kappa,\lambda)$ results in the ``Ricci'' tensor: \be
R^{\alpha\beta\mu\nu}=
\Gamma^{\alpha\beta\kappa\lambda,\mu\nu}_{\kappa\lambda}-
\Gamma^{\alpha\beta\mu\nu,\kappa\lambda}_{\kappa\lambda}+
\Gamma^{\alpha\beta\kappa\lambda}_{\omega\xi}
\Gamma^{\omega\xi\mu\nu}_{\kappa\lambda}-
\Gamma^{\alpha\beta\mu\nu}_{\omega\xi}
\Gamma^{\omega\xi\kappa\lambda}_{\kappa\lambda} \ee A lengthy but
straightforward calculation, gives: \be \label{superricci}
R^{\alpha\beta\mu\nu}=\frac{1}{8}(n\gamma^{\alpha\mu}\gamma^{\beta\nu}
+n\gamma^{\alpha\nu}\gamma^{\beta\mu}-2\gamma^{\alpha\beta}\gamma^{\mu\nu})
\ee With the help of (\ref{superricci}) and
(\ref{supercontravariantmetric}) the ``Ricci'' scalar is found to
be: \be \label{superricciscalar}
R=L_{\alpha\beta\mu\nu}R^{\alpha\beta\mu\nu}=\frac{1}{4}(n^{3}+n^{2}-2n)
\ee

Finally, the ``conformal Beltrami'' operator, is: \be
\label{superBeltrami} \Box^{2}_{c}\equiv
\Box^{2}+\frac{\mathcal{D}-2}{4(\mathcal{D}-1)}=L_{\alpha\beta\mu\nu}
\frac{\partial^{2}}{\partial
\gamma_{\alpha\beta}\gamma_{\mu\nu}}-L_{\alpha\beta\mu\nu}
\Gamma^{\alpha\beta\mu\nu}_{\kappa\lambda}
\frac{\partial}{\partial
\gamma_{\kappa\lambda}}+\frac{\mathcal{D}-2}{4(\mathcal{D}-1)}R
\ee where $\mathcal{D}$ is the dimension of the general metric
space: $\mathcal{D}=\frac{n(n+1)}{2}$, i.e. the number of the
independent
$\gamma_{\mu\nu}$'s.\\
One can find that: \be \label{superlinearterm}
L_{\alpha\beta\mu\nu}\Gamma^{\alpha\beta\mu\nu}_{\kappa\lambda}=
\frac{3-n^{2}}{n-1}\gamma_{\kappa\lambda} \ee thus
(\ref{superBeltrami}), takes the form: \be
\label{superfinalBeltrami} \Box^{2}_{c}\equiv
L_{\alpha\beta\mu\nu} \frac{\partial^{2}}{\partial
\gamma_{\alpha\beta}\gamma_{\mu\nu}}-\frac{3-n^{2}}{n-1}\gamma_{\kappa\lambda}
\frac{\partial}{\partial\gamma_{\kappa\lambda}}
+\frac{\mathcal{D}-2}{4(\mathcal{D}-1)}R \ee

\end{document}